\documentclass[a4paper,10pt]{aastex}

\title{Long Term X-ray Monitoring Of The TeV Binary LS I +61 303 with RXTE}

\author{A. Smith\altaffilmark{1}, P. Kaaret\altaffilmark{2}, J. Holder\altaffilmark{3}, A. Falcone\altaffilmark{4}, G. Maier\altaffilmark{5}, D. Pandel\altaffilmark{2}, M. Stroh\altaffilmark{4}}
\email{awsmith@anl.gov}
\altaffiltext{1}{Argonne National Laboratories, 9700 S. Cass Avenue, Argonne, IL 60439, USA }
\altaffiltext{2}{Department of Physics and Astronomy, University of Iowa, Van Allen Hall, Iowa City, IA 52242, USA}
\altaffiltext{3}{Department of Physics and Astronomy and the Bartol Research Institute, University of Delaware, Newark, DE 19716, USA}
\altaffiltext{4}{Department of Astronomy and Astrophysics, 525 Davey Lab.,Penn. State University, University Park, PA 16802, USA}
\altaffiltext{5}{Physics Department, McGill University, Montreal, QC H3A 2T8, Canada}

\begin{document}



\begin{abstract}
We report on the results of a long term X-ray monitoring campaign of the galactic binary LS I +61 303 performed by the Rossi X-ray Timing Explorer. This dataset consists of 1 ks pointings taken every other day between 2007 August 28 until 2008 February 2. The observations covered six full cycles of the 26.496 day binary period and constitute the largest continuous X-ray monitoring dataset on LS I +61 303 to date with this sensitivity. There is no statistically strong detection of modulation of flux or photon index with orbital phase; however, we do find a strong correlation between flux and photon index, with the spectrum becoming harder at higher fluxes. The dataset contains three large flaring episodes, the largest of these reaching a flux level of 7.2$^{+0.1}_{-0.2}$$\times$10$^{-11}$ erg cm$^{-2}$s$^{-1}$ in the 3-10 keV band, which is a factor of three times larger than flux levels typically seen in the system. Analysis of these flares shows the X-ray emission from LS I +61 303 changing by up to a factor of six over timescales of several hundred seconds as well as doubling times as fast as 2 seconds.  This is the fastest variability ever observed from LS I +61 303 at this wavelength and places constraints on the size of the X-ray emitting region.
\end{abstract}

\section{Background}

The galactic binary LS I +61 303 is one of the most heavily studied binary star systems in the Milky Way, and is one of only three so called ``TeV Binaries'' to be regularly detected in the $>$100 GeV gamma-ray regime; the other two being LS 5039 (Aharonian et al. 2005a) and PSR B1259-63 (Aharonian et al. 2005b). Although the subject of many observational campaigns, the fundamental identification of the components of the system remains relatively unclear. From observations detailed in Hutchings and Crampton (1981) and Casares et al. (2005) it is clear that the system can be classified as a high mass X-ray binary (HMXB) located at a distance of 2 kpc; the components of the system consisting of a compact object in a 26.496($\pm$0.003) day orbit around a massive BO Ve main sequence star. Although the range of masses derived for the compact object favor a neutron star component, a black hole cannot be ruled out (Casares et al. 2005). Periastron passage of the compact object occurs at $\phi=$0.23 (here $\phi$ represents the orbital phase ranging from 0.0 to 1, $\phi$=0.0 set at  JD 2443366.775), with apastron passage occuring at 0.73, and inferior and superior conjunctions occuring at 0.26 and 0.16, respectively (Casares et al 2005). 

The two main competing scenarios to explain the system can be summarized as $\textit{microquasar}$ (i.e. non-thermal emission powered by accretion and jet ejection) or $\textit{binary pulsar}$ (i.e. non-thermal emission powered by the interaction between the stellar and pulsar winds). The microquasar model (Massi et al. 2001,  Bosch-Ramon et al. 2006) used to describe this system has several drawbacks (for example, lack of blackbody X-ray spectra), however the scenario has not been ruled out, for example see Romero et al. (2007). There is growing evidence for the binary pulsar model of the system (Maraschi and Treves 1981, Dubus 2006), most strongly supported by the result of Dhawan (2006) in which VLBA monitoring of the system revealed a cometary radio structure around LS I +61 303 which was interpretated as the interaction between the pulsar and Be star wind structures. However, there is currently no evidence for the presence of a pulsar within the system either in radio or X-rays. 

LS I +61 303 has been historically an object of interest due to its quasi-periodic outbursts at radio (Paredes et al. 1998, Gregory 2002) and X-ray wavelengths (Taylor et al. 1996, Leahy et al. 1997, Paredes et al. 1997, Harrison et al. 2000, Greiner and Rau 2001). The radio outbursts are well correlated with the orbital phase (Gregory 2002),  although the phase of maximum emission can vary between 0.45 and 0.95. The first high energy association of the source was with the COS-B source 2CG 135 +01 (Hermsen et al. 1977). LSI +61 303 has also been identified with the EGRET source 3EG J0241+6103, a source which also shows evidence for a 26.5 day modulation in the GeV band (Massi 2004b). More recently, LS I +61 303 has been detected as a variable TeV gamma-ray source (Albert et al. 2006, Acciari et al. 2007) with high emission appearing near apastron.

The collection of X-ray observations on LS I +61 303 is expansive, however, the exact character of the X-ray emission from this source remains unclear. The X-ray source was originally weakly identified by the Einstein satellite (Bignami et al. 1981). Further observations of the system by ROSAT in 1991 and 1992 (Goldoni and Mereghetti 1995, Taylor et al. 1996) in the 0.01-2.4 keV band showed the source to be variable by a factor of 3 over a single orbital cycle; a peak flux occuring between orbital phases 0.4 and 0.6. The first accurate observations in the $>$5 keV range were performed by ASCA (Leahy et al. 1997) in the 0.5-10 keV band. ASCA performed two separate 30 ks exposures near orbital phases 0.2 and 0.45 showing relatively low flux states (0.6/0.4 $\times$10$^{-11}$erg cm$^{-2}$s$^{-1}$). The spectra from these observations were well fit by an absorbed powerlaw with photon indices of 1.7$\pm$0.1 and 1.8$\pm$0.1, respectively.  RXTE observed LS I +61 303 for an entire orbital cycle in 1996, seeing a clear increase in flux between $\phi$=0.3 and 0.7, with a peak occurring near phase 0.45 at a flux of 2 $\times$10$^{-11}$erg cm$^{-2}$s$^{-1}$ (Harrison et al. 2000). There have been three separate published analyses of this dataset (Harrison et al. 2000, Greiner and Rau 2001, Neronov and Chernyakova 2006). All three analyses find that a simple absorbed power law provides a resonable fit to the derived spectra. In 2002, XMM-\textit{Newton} was used to monitor LS I +61 303 in the 0.2-12 keV range (Neronov and Chernyakova 2006). Four 5-6 ks pointings were taken spread out over a single orbit, with a single 6 ks pointing taken several months later. These pointings showed the 2-10 keV flux to be highly variable, having a minimum near periastron and peaking near phase 0.55. The spectral fitting (again, using an absorbed powerlaw) was consistent with a photon index value of 1.5 for 4 out of 5 of the pointings, with a much softer index of 1.78 resulting for the data point just preceding the transition from a high to low X-ray flux. Neronov and Chernyakova (2006) interpreted this as evidence for correlation between spectral behavior and flux states. 

Several years later in 2005, XMM-\textit{Newton} was again used for an extended pointing (48.7 ks) on LS I +61 303 during orbital phase 0.61 (Sidoli et al. 2006). This observation showed evidence for variation of the hardness ratio (defined as the ratio between 0.3$\rightarrow$2 keV and 2$\rightarrow$12 keV emission) over the timescale of hours. These observations also showed a sharp decrease in flux of the order of a factor of three over a time period of a few thousand seconds, with flux decreasing from 1.2$\rightarrow$0.4 $\times$10$^{-11}$erg s$^{-1}$cm$^{-1}$, which were the first observations to detail such rapid variability of the X-ray flux from this system. The Chandra satellite carried out a 50 ks pointed observation of LS I +61 303 in April 2006 (Paredes et al. 2007) near orbital phase 0.04. These observations (0.3 to 10 keV) revealed the presence of kilosecond scale miniflares in the source, with emission increasing by a factor of 2 over a timespan of roughly one hour. Similar to the XMM-\textit{Newton} extended exposure, the Chandra observations also show an implied correlation between harder emission and increased flux. The mean flux was determined to be 0.71$^{+0.18}_{-0.14}$ $\times$ 10$^{-11}$erg s$^{-1}$cm$^{-1}$ which is also consistent with previous measurements near that orbital phase. The derived photon index, however, was determined at $\alpha$=1.25$\pm$0.09, much harder than any previous measurements made. The Chandra exposure also found evidence (3.2$\sigma$ significance) for extended emission between 5 and 12.5" towards the North of LS I +61 303, an indication that particle acceleration resulting in X-ray emission may be taking place as far away as 0.05-0.12 parsecs from LS I +61 303. 

Taken as a whole, the X-ray observations conducted on LS I +61 303 indicate that the source is an extremely unpredictable one. Although long term X-ray monitoring conducted by the All Sky Monitor aboard RXTE indicate that the system exhibits a long term 26.42 $\pm$0.05 day period in X-ray emission (Leahy 2001), the X-ray observations  descibed above show that extreme variation in the lightcurve is seen between different orbital cycles. As for spectral behavior it is clear that a simple absorbed powerlaw provides an effective fit to the observational data, however, there is mounting evidence that the photon index may be correlated to the flux level of the system. 

\section{RXTE Observations and Data Reduction}

Between August 28 2007 and February 2 2008 (MJD 54340-54504) a total of $\sim$80ks of observation time was accumulated with the Proportional Counting Array (PCA) instrument aboard RXTE (Jahoda 2006). 1 ks exposures were taken every other day which resulted in good observational coverage of a total of six 26.5 day orbital cycles of the binary system. Two separate data modes were used in this analysis: for lightcurve and spectral analysis the ``Standard 2'' mode was used (16s accumulation time, 129 energy channels), for analysis requiring greater time resolution the ``Good Xenon'' mode was utilized (1$\mu$s resolution, 256 energy channels). Data in both modes were  reduced using NASA HEASARC's FTOOLS 6.5 package (Blackburn 1995). Data were selected using standard quality criteria cuts as suggested by the RXTE Guest Observer Facility data reduction page\footnote[1]{$ http://heasarc.gsfc.nasa.gov/docs/xte/data\_analysis.html$} with the exception of the use of a slightly looser restriction on the suggested ``Time Since South Atlantic Anomaly Passage''  (25 minutes as opposed to the suggested 30 minutes). It was determined that the quality of the data was not effected by this relaxation and the exposure time was significantly increased for a significant percentage of the pointings.

 During the time that this dataset was accumulated, the number of PCA units which were activated changed per observation, therefore, for the Standard 2 analysis, a spectrum from 3-10 keV was extracted from each nights individual PCA configuration utilizing combinations of PCA units 2, 3 and 4 (since PCUs 0 and 1 have lost their propane layer). The rationale behind this was that LS I +61 303 is typically not a stong X-ray emitter above this energy range and due to RXTE instrument degradation, the energy channels below 3 keV are not reliable (Jahoda 2006) at this point in time. Only the top PCU layer was utilized in order to maximize signal to noise. After generating simulated background spectra using $\emph{pcabackest}$ and creating PCA response matrices with $\emph{pcarsp}$, XSPEC12 was used to fit each night's background subtracted spectra. Previous observations of LS I +61 303 with RXTE (Harrison et al. 2000) as well as XMM-$\textit{Newton}$ (Sidoli et al. 2006) and Chandra (Paredes et al. 2007) found that the best fit to the observed spectra was provided by a simple, absorbed power-law (without a blackbody component) and this is the approach that we adopt in the analysis presented here. In this fitting, as with previous RXTE analyses of this source, the value of the galactic N$_{H}$ was kept constant; fixing the absorbing hydrogen column density at (N$_{H}$) to 0.75$\times10^{22}$cm$^{-2}$ (Kalberla et al. 2005). Therefore, the spectral fit used here takes the form of:
\begin{equation}
A(E) = Ke^{-N_{H}\sigma(E)}\left(\frac{E}{keV}\right) ^{-\alpha}
\end{equation}
where K is a normalization constant at 1 keV, $\sigma$ is the photo-electric cross-section, and $\alpha$ is the photon index. The $\chi^{2}$/d.o.f. values for these fits were acceptable, with all values falling between 0.25 and 1.2 for 14 degrees of freedom. Keeping with standard convention, the flux values were then generated by integrating the best fit spectra from 2-10 keV along with the associated 1$\sigma$ error levels, which are used throughout this work. When producing Standard 2 data combining multiple observations (and PCA configurations) such as the phase and flux binned datasets, it was neccessary to produce spectra (and fluxes) in a slightly different manner. Since combining large spectral datasets with differing PCA configurations (with associated differing calibrations) can produce large systematic errors, it is preferrable to only use a single common PCU for all available datasets.  Since PCU 2 was active for all observations in the current dataset, spectra from PCU 2 were produced and combined to give a single spectrum for each bin in question. 

For Good Xenon data ($<$16s time resolution), the same overall quality selection criteria were utilized to produce lightcurves from PCU 2 only. Background subtraction for these lightcurves was carried out by using the FTOOLS routine \textit{lcmath} to subtract an estimated background lightcurve (produced from Standard 2 data) from the Good Xenon observation as recommended by the RXTE guest observer facility.

\section{Results}

\subsection{Overall Results}
Shown in Figure 1 is the overall lightcurve in daily bins along with 1$\sigma$ error bars. As can be seen, the flux typically modulates between 0.5-2 $\times$ 10$^{-11}$erg s$^{-1}$cm$^{-2}$ over the timespan of the 26.5 day orbital cycle. The average flux presented was 1.67$\pm$0.02 $\times$ 10$^{-11}$erg s$^{-1}$cm$^{-2}$, with a reduced $\chi^{2}$ value of 30.6 (75 d.o.f.) for a constant flux fit; demonstrating strong variaibility over the span of the dataset ($<$0.01$\%$ probability of being constant). The most immediately obvious feature of the lightcurve is the presence of three exceptionally large flares (FL1, FL2, FL3) within the data appearing on September 13, 15, and 29 2007 (MJD 54356, 54358, 54372) presenting flux levels of 7.2($^{+0.1}_{-0.2}$), 3.5($^{+0.1}_{-0.2}$), and 4.9($^{+0.1}_{-0.2}$)$\times$ 10$^{-11}$erg s$^{-1}$cm$^{-2}$ respectively.  These powerful flares represent the most extreme X-ray activity detected from LS I +61 303 to date, with the largest of these flares, FL1,  presenting an inferred luminosity (at a distance of 2 kpc) of 3.4$\times$10$^{34}$erg/s. These flares will be discussed in greater detail below.

 To examine the overall structure of the RXTE data with respect to the 26.5 day orbital cycle of the system, the individual observations were placed in 0.1 $\phi$ bins and analyzed together as described in section 2. The extracted 2-10 keV flux and fitted spectral indices from PCU 2 for these bins are shown in Figure 2 (left) and in Table 1. A constant flux fit to the data gives a reduced $\chi^{2}$/d.of. value of 33.68 (9 d.o.f.) indicating that the flux value has a $<$0.01$\%$ probability of being constant over the orbital phase.

When examining the orbital phase versus photon index behavior on the right side of Figure 2, it is not clear that the index changes significantly over the orbital phase. Although it appears that the spectrum gets softer between orbital phases 0.5$\rightarrow$0.7, a constant index fit to this lightcurve gives a reduced $\chi^{2}$ value of 2.58 (9 d.o.f.), indicating a $<$5.5$\%$ probability of being constant over the orbit. 

 The harder spectral indices of the flaring episodes motivated a search for a possible correlation between flux and photon index. To carry this out, the data were binned by flux levels and re-analyzed by extracting a single PCU 2 spectrum from all data falling within the flux windows of F$<$1.0, 1.0$<$F$<$1.5, 1.5$<$F$<$2.0, 2.0$<$F$<$2.5, and F$>$2.5 $\times$ 10$^{-11}$erg s$^{-1}$cm$^{-1}$ (see Table 2 and Figure 3 left). As can be seen there is a clear correlation between the flux level and photon index, with the higher emission states resulting in a harder photon index. This relationship has a Pearson product-moment correlation coefficient of -0.9867 at $>$99$\%$ confidence level. To provide a check that this result is not biased by the process used of extracting the flux from the fitted spectra (i.e. a harder spectrum will naturally result in a higher flux for a given normalization), the count rate per second for PCU 2 only was extracted for each of the flux bins used (see Figure 3 right). The Pearson product-moment correlation between count rate and photon index is -0.966 at  $>$99$\%$ confidence level.

\subsection{FL1, FL2, FL3}
As with the overall lightcurve shown in Figure 1, the data covering the three flaring episodes were taken with varying combinations of PCUs which were combined and fit with a single spectrum. For each flare, the count rates from all PCU units were compared in order to ensure that the flaring behavior was not due to spurious electrical discharging in a single PCU unit. All PCU units showed similar count rates for the flare data, indicating that the flaring behavior was due to source flux and not instrumental error. It is also worth noting that none of the flare data presented here were taken near an SAA passage, therefore contamination due to this can be ruled out. The specific properties of each flare can be seen in detail in Table 3.

FL1 (MJD 54356) stands as the most powerful of the three flares observed within this dataset as well as the most powerful detected from this source to date, presenting a flux of 7.2$\pm$0.2 $\times$ 10$^{-11}$erg s$^{-1}$cm$^{-2}$.  When analyzed in Good Xenon event mode, the data has a sufficient count per second rate to be binned in 2s intervals (see Figure 4). As can be seen, the flux varies by up to a factor of 6 within the tens of seconds timescale. To illustrate this point and investigate the fastest doubling time within the flare, the first 100 seconds of the flare are zoomed in upon in Figure 4 (right). Directly after both t= 20s and 60s the count rate increases by more than a factor of two within the bin time of the data points, placing a flux doubling time limit of $<$2s on the X-ray emission from this source. 

FL2 (MJD 54358) was observed at a flux level of 3.5$^{+0.1}_{-0.2}$$\times$ 10$^{-11}$erg s$^{-1}$cm$^{-2}$. This flux level is not as high as FL1, but it is still large enough to merit closer investigation. Since the count rate is sufficiently smaller than that in FL1, the Good Xenon data shown in Figure 5 is binned in 5 second intervals. The data does not show as large variations as FL1, however it does contain several well defined flares which rise and fall on the 10-20s timescale. One of these episodes is shown in Figure 5 (right).

FL3 (MJD 54372) was observed at a flux level of 4.9$^{+0.1}_{-0.2}$$\times$ 10$^{-11}$erg s$^{-1}$cm$^{-2}$. The flux levels are nearly as chaotic as those seeen in FL1, with the flux varying by nearly a factor of six over the 100 second timescale (see Figure 6 left). The zoomed region to the right of Figure 6 shows another example of a rapid flare within the data.

To search for periodic signals in these flares, a timing analysis was carried out in the following manner: the data from the Good Xenon mode was re-extracted from PCU 2 with no quality selection criteria (in order to maximize the temporal length of the testable dataset). Although the removal of filtering criteria carries a risk of increasing the noise level in the data and therefore altering the result of a periodicity search, the data was examined both with and without filtering criteria and it was determined that the contribution of increased noise (for example electron noise in the detector) was negligible compared to the flux levels observed from the source (determined by the data examined with filtering criteria applied). The data was background subtracted using the same method as above and barycentered using the ftools routine $\textit{faxbary}$. The relaxation of the quality selection criteria led to a substantial increase of the testable data set of $\sim$400s for each flaring episode. The data was tested using the ftools FFT routine $\textit{powspec}$. The data was binned into 5 ms bins and a power spectrum constructed between 0.003 Hz and 100 Hz.  Both FL1 and FL3 showed a strong red noise component (i.e. increasing power at lower frequencies) in the test results due to their flux generally decreasing over the observation interval. To remove this component, a first order polynomial was fit to the data and the best fit result was subtracted from the input lightcurve to be tested. It is these ``de-trended'' power spectra which are shown in Figure 7. The powers shown in Figure 7 are normalized by dividing the resulting power by the number of bins in each tested interval.

For FL1, the relaxation of the quality selection resulted in the observation length increasing to 896s. The results of the timing analysis are shown in Figure 7. There are no statistically significant structures evident in the power spectrum.  FL2 was increased to 912s by the relaxation of the selection criteria, with no significant structures appearing in the timing analysis either (Figure 8). FL3 was increased to 904s when re-extracted, and the results of its timing analysis are shown in Figure 9. While there are no high frequency features present within the power spectrum,  there is an excess near 0.003 in the low frequency end corresponding to a period of $\sim$330s. However, since the light curve duration is only three times the inferred period, identification of this feature as periodic or quasiperiodic is not warranted.

\section{Summary}

The current RXTE dataset demonstrates the overall variability of the X-ray behavior of the system. Additionally, we find a strong correlation between photon index and flux from the system. While previously believed to be an X-ray source with somewhat dependable modulation, the observations reported here show strong flares which complicate the measurement of the typically expected flux and spectral variations as a function of binary orbital phase. The X-ray behavior (with respect to the orbital phase) is not completely understood.

The three large flares observed in this dataset represent the strongest X-ray activity recorded from this system to date with luminosities exceeding 3$\times$10$^{34}$erg s$^{-1}$. These flares show that the flux levels within the flare structures rise and fall by factors of up to 6 on the 100s time scale, with doubling times as short as 2 seconds (FL1). Through this variability, the causal size of the emission region can be constrained to be less than R$<c\Delta$t = 6$\times$10$^{10}$ cm. It should be noted that this constraint does not account for the scenario in which the emission is originating from a relativistic jet with a Doppler factor which would modify the above calculation. Since the existence of such a jet within the system is not clearly evident (and much less its Doppler factor), this modification is neglected at this time. Although rapid flaring over the kilosecond timescale has been demonstrated in both Chandra and XMM-$\textit{Newton}$ data, this is the first indication that the X-ray flux from the sytem varies on such a rapid scale.

We note that fast X-ray flaring behavior has been observed in binary X-ray pulsars such as IGR J16465-4507 and AX J1841.0-0536 (Lutivinov et al. 2005, Sguera et al. 2006 ). Following along these lines,  a possible scenario which could be used to explain the available observations is the binary pulsar model of Zdziarski et al. (2008). In this scenario, the radio through TeV emission results from the interactions between the pulsar wind and a two-component Be star wind. The two components consist of a fast ``clumpy'' polar wind and a dense circumstellar disk. When the pulsar is close enough to the Be star to be heavily exposed to the fast polar wind, clumps of relativistic electrons in the polar wind are maintained by magnetic field inhomogeneities and are sufficiently energetic to cause mixing with the pulsar wind. Shocks between the clumps and the pulsar wind accelerate electrons which can then inverse-Compton (IC) scatter off of the Be star photons, giving rise to the highly variable X-ray emission. Clumps such as these have been observed on the scale of 10$^{11}$ cm in 2S 0114+65 (Apparao, Bisht, and Singh 1991), which agrees (within error) with the implied timescale for variability observed in this work. However, even if clumps are much larger than this, it is believed that the interaction between a clump and the pulsar wind will cause a flare in a region with characteristic size equal to the stand-off distance between the pulsar and equatorial stellar wind (Dubus 2008b). At periastron passage the stand-off distance can be as small as 5$\times$10$^{10}$ cm, which is consistent with the variability reported in this work. On a larger timescale, the pulsar wind moves through region of varying density of the equatorial wind. During traversal of the densest parts of the disk, X-ray emission is quenched due to Coulomb losses, and when the pulsar moves towards the more spare regions of the disk, IC losses can dominate over both Coulomb and synchrotron mechanisms, giving rise to a orbital modulation of the X-ray flux. Although this dataset does not strongly indicate a modulation of the X-ray flux with orbital phase (possibly due to complications from shorter timescale flares), the possibility of low amplitude modulation is still consistent with the observations reported here. 

 We also note that fast X-ray variability such as that reported here has been observed in good candidates for galactic black hole systems such as Cygnus X-1 (Gierlinski and Zdziarski 2003) and GRS 1915 +105 (Belloni et al. 2000); which are both known accretion driven sources. RXTE observations of the TeV binary LS 5039 (Bosch-Ramon et al. 2005a) show a photon index versus flux correlation and hourly timescale flux variations similar to what we find for LS I +61 303, but no very rapid variability. The authors interpret these results in an accretion driven scenario where a stellar wind feeds a disk around either a black hole or neutron star and argue that the X-ray emission is caused by a population of relativistic electrons accelerated within the jet, which is in turn fed by the wind fed accretion disk. These electrons are responsible for the observed X-ray emission via either the synchrotron process or inverse-Compton scattering of local stellar photons. In this scenario, fast variations in the stellar wind are responsible for the observed ``miniflares'', while sudden increases in accretion (higher flux) would also result in a higher electron acceleration efficiency in the jets. This efficiency increase results in a higher peak of the electron energy distribution,  in turn resulting in a harder emission spectra. Additionally, in the model described in Bosch-Ramon et al. (2005b),  they predict a jet length of $\sim$10$^{11}$ cm, which is consistent with the several second variability described in this work. It should be noted that no X-ray spectral signatures of an accretion disk (i.e. a blackbody spectral component) are found in LS 5039 either, further reinforcing the link between the X-ray emission of the two sources. 

Recently, Swift-BAT has reported the detection of a short (0.23s), extremely powerful X-ray burst with a luminosity of 10$^{37}$ erg in the 15-150 keV energy range within the 90$\%$ containment radius of LS I +61 303 (Barthelmy et al. 2008). Follow up observations with the Swift-XRT instrument 921 seconds later showed no significantly elevated flux in the 0.3-10 keV energy range (0.92$\times$10$^{-11}$ erg cm$^{-2}$s$^{-1}$). While the authors of this report note the possibility that this emission was due to an unrelated short gamma-ray burst, they claim that the evidence is in favor of activity from a source within LS I +61 303. Dubus and Giebels (2008) submit that this burst could be evidence in favor of magnetar-like activity from a source within LS I +61 303. Although reconciling the presence of a magnetar within LS I +61 303 with previous X-ray observations is somewhat problematic (Bosch-Ramon 2008, Dubus 2008a), this could be the first magnetar to be discovered within a high mass X-ray binary system. If confirmed with following observations, this type of extremely powerful bursting may resolve the question of the identification of LS I +61 303.

In conclusion, while neither conclusively ruling out or confirming either the microquasar or binary pulsar scenarios, the observations reported here will aid further modeling work of LS I +61 303 by the demonstrated constraint on emission region size, as well as the correlation between flux and photon index.

\acknowledgements
The authors wish to thank the anonymous referee for their timely and helpful comments. We also wish to thank Valenti Bosch-Ramon, Guillaume Dubus, Marc Ribo, and Masha Chernyakova for providing helpful comments and suggestions on this work.
{\it Facilities:} \facility{RXTE}

\begin{figure}
\begin{center}
   \includegraphics[width=\textwidth,height=100mm]{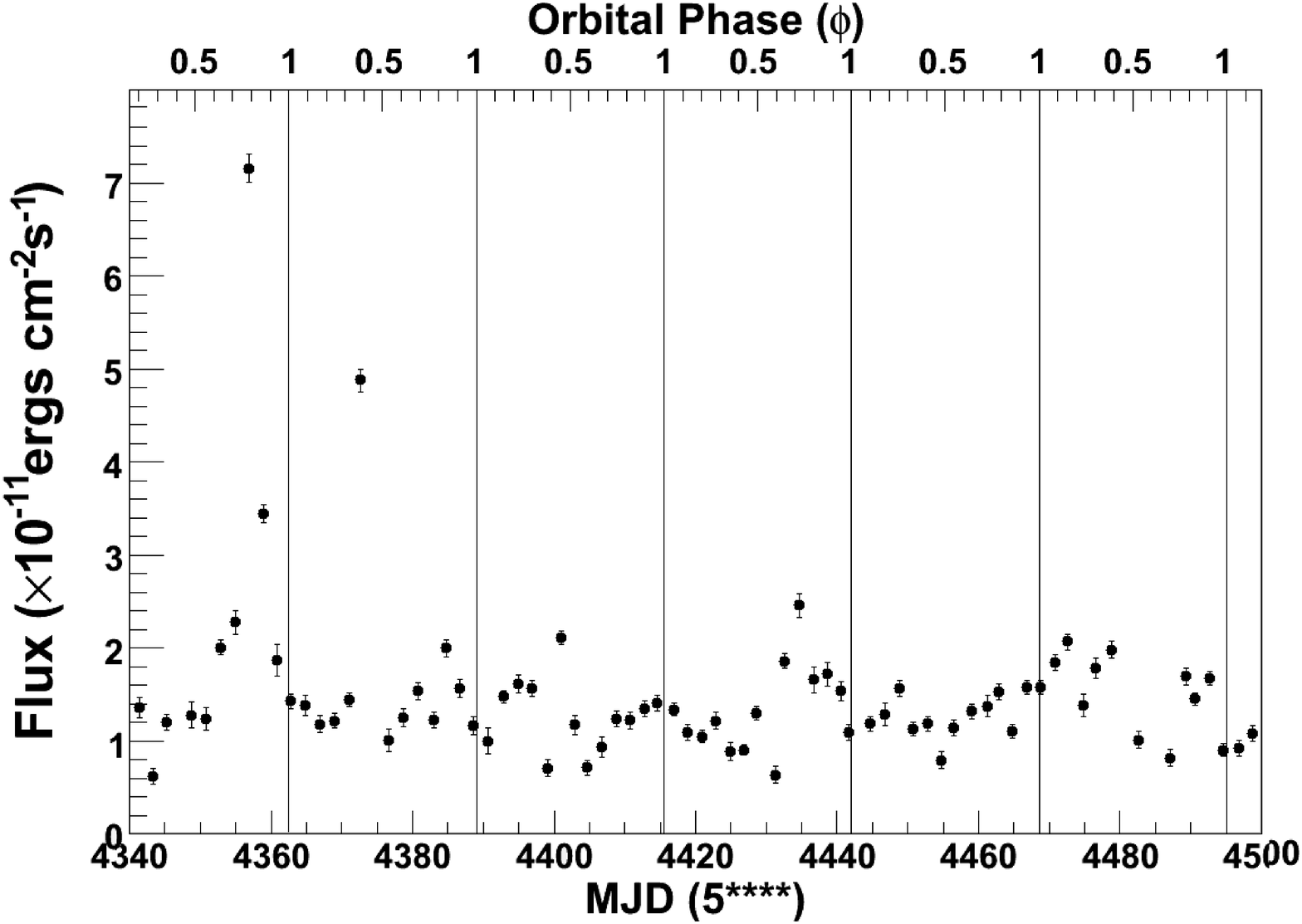}

\end{center}
\caption{The 2007/8 RXTE campaign on LS I +61 303. Fluxes shown are for X-rays in the 2-10 keV energy band, with 1$\sigma$ error bars.}
\end{figure}

\begin{figure}
\begin{center}
   \includegraphics[width=\textwidth,height=60mm]{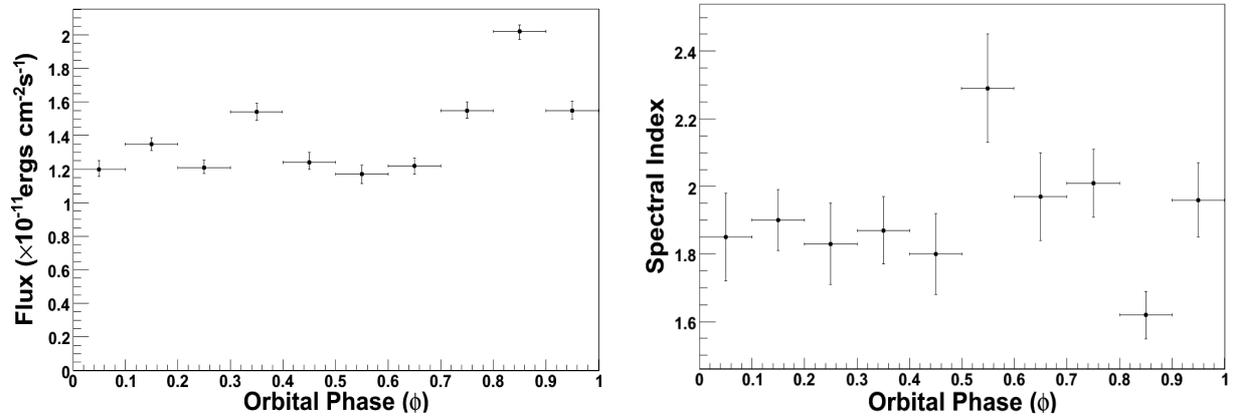}

\end{center}
\caption{The total RXTE dataset reanalyzed in phase bins of 0.1 $\phi$. The left figure shows the 2-10 keV flux versus orbital phase and the right figure shows the photon index as a function of orbital phase. }
\end{figure}

\begin{figure}
\begin{center}
   \includegraphics[width=\textwidth,height=60mm]{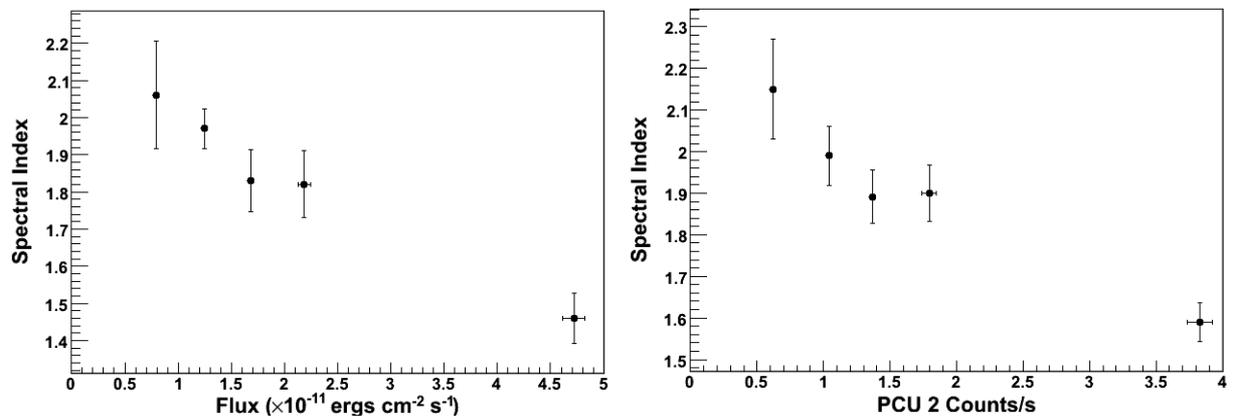}

\end{center}
\caption{The 2-10 keV flux versus the associated photon index from the flux bins described in the text. A clear correlation between higher flux and harder spectral behavior. On the right is shown the PCU2 counts from the same bins compared to the photon index. }
\end{figure}

\begin{figure}
\begin{center}
   \includegraphics[width=\textwidth,height=60mm]{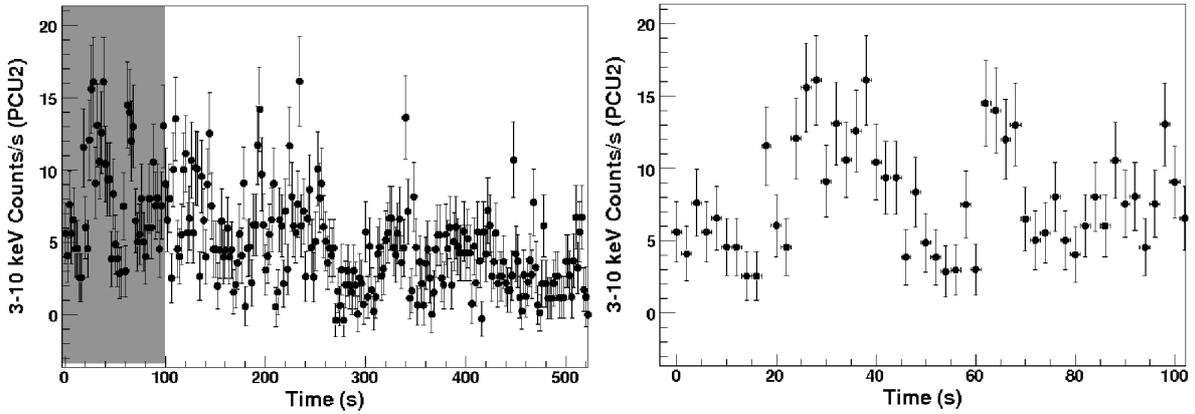}

\end{center}
\caption{Flare 1 binned in 2 second bins (left) with t=0 representing the beginning of the observation window. The flux increases by over a factor of 6 during 10-100s timescales. The shaded grey region is expanded in the right figure showing the source count rate doubling in less than the 2 second bin time.}
\end{figure}
\begin{figure}
\begin{center}
   \includegraphics[width=\textwidth,height=60mm]{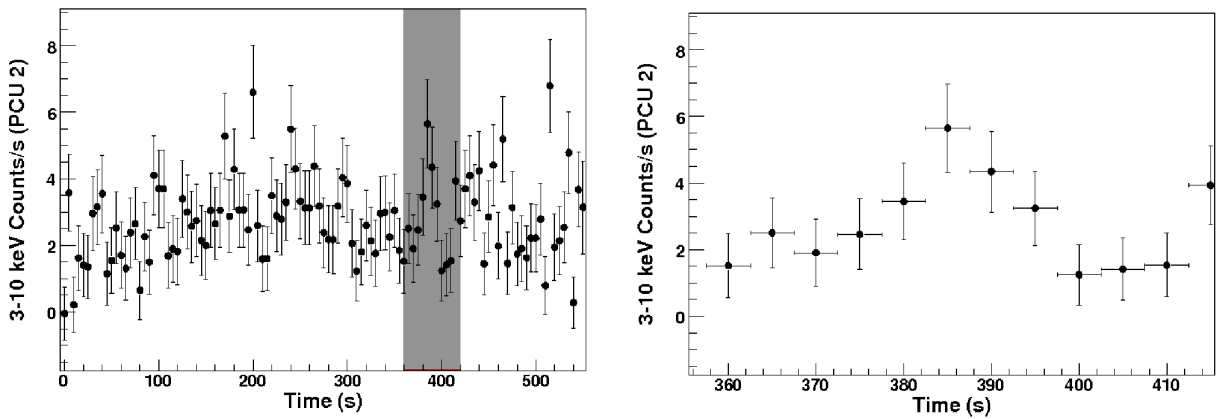}
\end{center}
\caption{Flare 2 binned in 5 second bins (left). The shaded grey region is expanded in the right figure which shows a well defined flare, doubling in flux, ocurring over the 10-20s timescale.}
\end{figure}
\begin{figure}
\begin{center}
   \includegraphics[width=\textwidth,height=60mm]{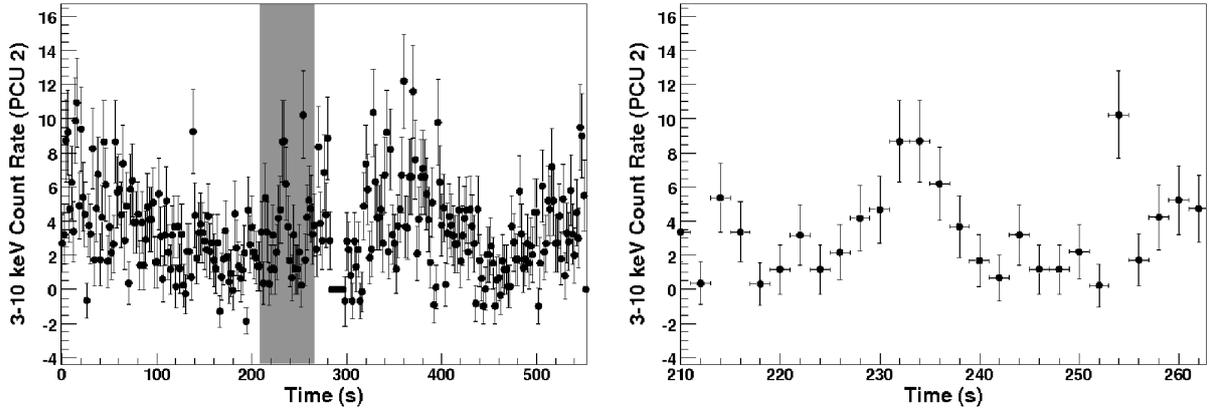}
\end{center}
\caption{Flare 3 binned in 5 second bins (left) along with the grey region expanded upon  (right) showing a flux doubling episode occurring over short timescales.}
\end{figure}

\begin{figure}
\begin{center}
   \includegraphics[width=\textwidth,height=120mm]{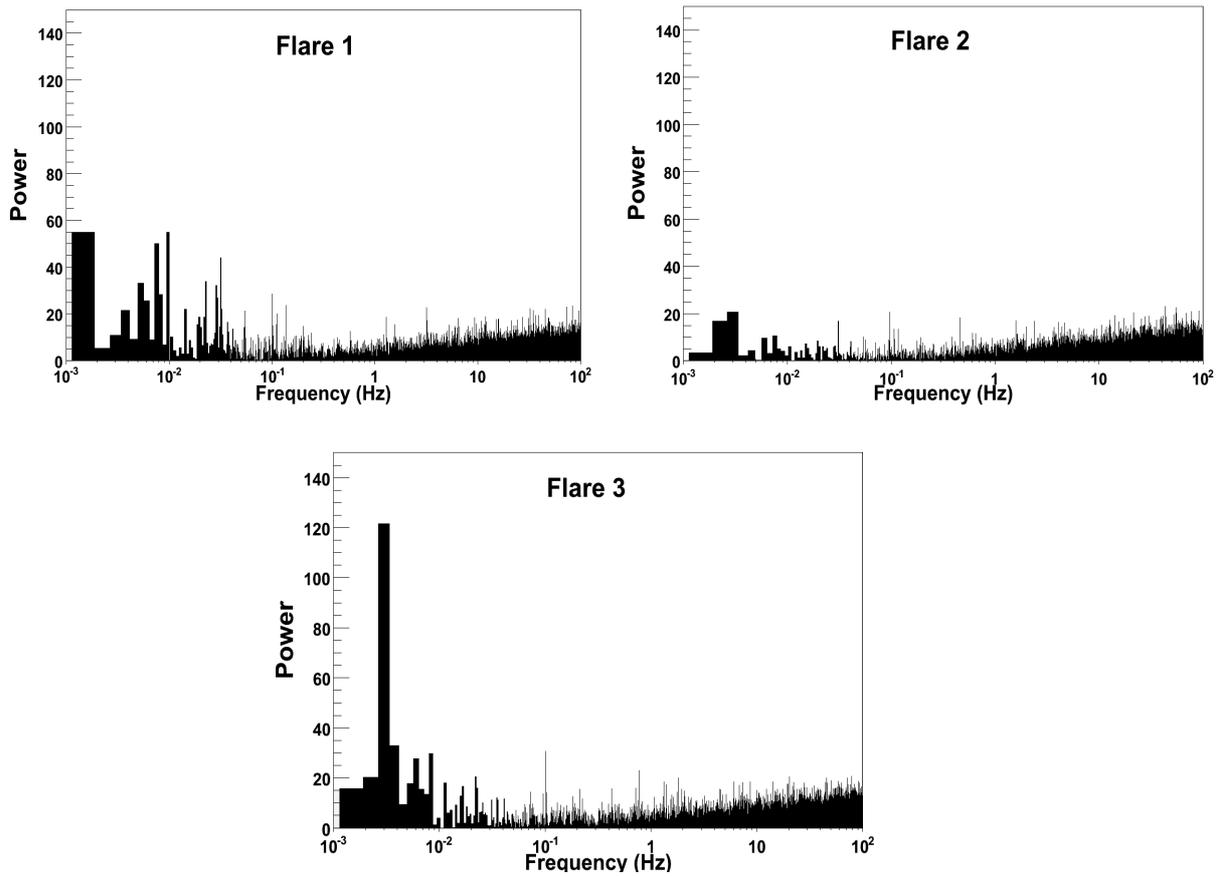}

\end{center}
\caption{The timing analysis of FL1, FL2, and FL3 as described in the text. No statistically significant features are found.}
\end{figure}

\begin{table}
\begin{center}
 
 \begin{tabular}{c|c|c|c}
\textbf{Orbital Phase} & \textbf{$\alpha$} & \textbf{Flux (2-10 keV)} & \textbf{$\chi^{2}$/d.o.f.}\\
\textbf{($\phi$)}&&\textbf{10$^{-11}$erg cm$^{-2}$ s$^{-1}$ }&    \\ \hline
0.0$\rightarrow$0.1&1.85$\pm$0.13&1.20$^{+0.05}_{-0.05}$  & {47.7}/{14} \\\hline 
0.1$\rightarrow$0.2 &1.90$\pm$0.09&1.35$^{+0.04}_{-0.04}$ & {5.5}/{14}    \\\hline
0.2$\rightarrow$0.3 &1.83$\pm$0.12&1.21$^{+0.04}_{-0.04}$ & {4.9}/{14}   \\\hline
0.3$\rightarrow$0.4 &1.87$\pm$0.10&1.54$^{+0.05}_{-0.05}$& {7.9}/{14}   \\\hline
0.4$\rightarrow$0.5 &1.80$\pm$0.12&1.24$^{+0.05}_{-0.05}$& {6.7}/{14}   \\\hline
0.5$\rightarrow$0.6 &2.29$\pm$0.16&1.17$^{+0.05}_{-0.05}$ & {11.6}/{14}   \\\hline
0.6$\rightarrow$0.7 &1.97$\pm$0.13&1.22$^{+0.05}_{-0.05}$ & {4.06}/{14}   \\\hline
0.7$\rightarrow$0.8 &2.01$\pm$0.10&1.55$^{+0.05}_{-0.05}$& {6.8}/{14}   \\\hline
0.8$\rightarrow$0.9 &1.62$\pm$0.07&2.02$^{+0.05}_{-0.05}$& {5.2}/{14}   \\\hline
0.9$\rightarrow$1.0 &1.96$\pm$0.11&1.55$^{+0.05}_{-0.05}$&  {12.4}/{14}  \\\hline

\end{tabular}
\end{center}
\caption{The results of the full dataset analysis, with the data binned into 0.1$\phi$ bins.}
\end{table}

\begin{table}
 \begin{tabular}{c|c|c|c}
\textbf{Flux Bin}&  \textbf{$\alpha$} & \textbf{ Average Flux (2-10 keV)} &\textbf{PCU 2 Count Rate}\\
\textbf{10$^{-11}$erg cm$^{-2}$ s$^{-1}$} &&\textbf{(10$^{-11}$erg cm$^{-2}$ s$^{-1}$})& \textbf{(Counts s$^{-1}$)}\\ \hline
F$<$10&2.06$\pm$0.15&0.79$^{+0.04}_{-0.04}$&0.62$\pm$0.03\\ \hline
10$<$F$<$15&1.97$\pm$0.05&1.25$^{+0.02}_{-0.02}$&1.04$\pm$0.02\\ \hline
15$<$F$<$20&1.83$\pm$0.08&1.69$^{+0.03}_{-0.03}$&1.37$\pm$0.02\\ \hline
20$<$F$<$25&1.82$\pm$0.09&2.19$^{+0.06}_{-0.06}$&1.80$\pm$0.05\\ \hline
F$>$25&1.46$\pm$0.07&4.72$^{+0.1}_{-0.1}$&3.83$\pm$0.09\\ \hline

\end{tabular}
\caption{The results of the flux versus photon index analysis with the data binned as described in the text.}
\end{table}

\begin{table}
\begin{center}

 \begin{tabular}{c|c|c|c|c|c}
 \textbf{Flare} &\textbf{MJD (UTC)}&  \textbf{Usable Data (s)} & \textbf{PCUs}  & \textbf{$\alpha$} & \textbf{Flux (2-10 keV)} \\ 
&&&&&\textbf{(10$^{-11}$erg cm$^{-2}$ s$^{-1}$)}\\\hline
Flare 1 &54356 (11:09-11:25) & 528 & 2,4 & 1.4$\pm$0.1 & 7.2$\pm$0.2 \\ \hline
Flare 2 & 54358 (10:13-10:28) &560 & 2,3,4  & 1.7$\pm$0.1 & 3.5$\pm$0.1 \\ \hline
Flare 3 &54372 (03:31-03:35) &553 & 2,4  & 1.6$\pm$0.1 & 4.9$\pm$0.1 \\\hline
 \end{tabular}
\end{center}
\caption{The properties of the observed flaring episodes FL1, FL2, and FL3 as described in the text. Shown are the MJD and UTC of the observation, the usable window of observation after quality cuts, the number of PCUs operating during the observation, and the fitted photon index and integrated flux of the observation.}
\end{table}

\end{document}